\documentclass[aps,prl,superscriptaddress,twocolumn,showpacs]{revtex4-1}

\usepackage[T2A]{fontenc}
\usepackage[cp1251]{inputenc}
\usepackage{graphicx}
\usepackage{amsmath}
\usepackage{caption3}
\usepackage{floatflt}
\usepackage{dsfont}
\usepackage{mathrsfs}
\usepackage{color}
\usepackage{amssymb}
\usepackage{epsfig}
\usepackage{wrapfig,lipsum,booktabs}

\begin{document}

\title{Bose condensation of direct excitons in an off-resonant cavity at elevated temperatures}
\author{\firstname{N.~S.}~\surname{Voronova}}
\email{nsvoronova@mephi.ru}
\affiliation{%
National Research Nuclear University MEPhI (Moscow Engineering Physics Institute), 115409 Moscow, Russia}
\affiliation{Russian Quantum Center, 143025 Skolkovo, Moscow region, Russia}
\author{\firstname{I.~L.}~\surname{Kurbakov}}
\affiliation{Institute for Spectroscopy RAS, 142190 Troitsk, Moscow, Russia}
\author{\firstname{Yu.~E.}~\surname{Lozovik}}
\email{lozovik@isan.troitsk.ru}
\affiliation{Institute for Spectroscopy RAS, 142190 Troitsk, Moscow, Russia}
\affiliation{MIEM, National Research University Higher School of Economics, 101000 Moscow, Russia}

\begin{abstract}
We propose a way to increase the lifetime of two-dimensional {\it direct} excitons and show the possibility to observe their macroscopically coherent state at high temperatures.
For a single GaAs quantum well embedded in photonic layered heterostructures with subwavelength period, we predict the exciton radiative decay to be strongly suppressed. Quantum hydrodynamic approach is used to study the Berezinskii-Kosterlitz-Thouless crossover in a finite exciton system with intermediate densities. As the system is cooled down below the estimated critical temperatures, the drastic growth of the correlation length is shown to be accompanied by a manyfold increase of the photoluminescence intensity.
\end{abstract}

\maketitle

\textit{Introduction.}---Despite long-standing theoretical predictions \cite{keldysh,lozovik,rice}, the experimental observation of a macroscopically coherent state of excitons --- bound pairs of electrons and holes in a semiconductor --- for decades remained a challenging task and a subject of heated discussions \cite{exc2,exc1,exc3,butov1,timofeev1,jetpl0840329,butov2_}. Exciton Bose-Einstein condensation (BEC), once realized, could provide a plethora of beautiful observable phenomena with excitons, such as stimulated backscattering and multi-photon coherence \cite{jetpl0740288}, topological effects \cite{prb091161413}, supersolidity \cite{prb082014508}, ballistic transport \cite{prb088195309}, spin vortices \cite{prb089035302} and currents \cite{lozovik,prl110246403}, {\it etc}. One of the major obstacles on this way, together with the inhomogeneities and excess of free carriers \cite{kukushkin}, is the relatively high exciton radiative recombination rate which hinders effective thermalization. Therefore attempts to experimentally achieve excitonic BEC were mostly focused on electronically engineered systems utilizing indirect excitons (IX) \cite{lozovik} in coupled quantum wells (CQWs) under the influence of electric field \cite{exc1,exc3,butov1,timofeev1,butov2_}, which allows lifetimes longer than the characteristic timescales of relaxation. Compared to IX, direct excitons have lifetimes too short for effective cooling, and they recombine before reaching the condensed state. However, as they are more tightly bound and allow much higher densities, direct excitons would offer notably higher critical temperatures of BEC.

While electronic engineering as described above was the most fruitful approach so far  elongating the excitons' lifetime and leading to coherence, there is yet another way to control their radiative properties, which utilizes {\it photonic} engineering. As well known from cavity quantum electrodynamics, if an excited light source is embedded into a photonic material environment, its recombination can be both greatly enhanced \cite{purcell} or inhibited \cite{kleppner}. It can be vastly employed in devices which performance is limited by spontaneous emission, such as low-threshold lasers, heterojunction transistors, single photon emitters, {\it etc}. For example, ways to experimentally control the spontaneous emission rate have been successfully demonstrated for quantum dots (QDs) in laterally structured microcavities \cite{bayer}, as well as quantum wells (QWs) and QDs in two-dimensional photonic crystals \cite{noda}.

In this paper, we show the possibility to suppress direct exciton recombination in a single QW by embedding it into an off-resonant cavity. Relying on our theoretical calculations, we propose two specific GaAs-based geometries for experimental realization, provide optical properties of these structures, and predict exciton lifetimes up to tens of nanoseconds and more. A hydrodynamic quantum field theory \cite{Popov_,kane_,minnhagen_,giorgini_} joined with the Bogoliubov description is then employed to investigate the Berezinskii-Kosterlitz-Thouless (BKT) crossover in the exciton gas in the regime of intermediate correlations (at elevated densities), allowing to estimate the critical temperatures of condensation and calculate the change in intensity of photoluminescence (PL) along normal direction as the system is cooled down below the crossover temperature.

\textit{Optical properties.}---Considering a single GaAs QW embedded in a periodic photonic heterostructure, we seek the gap opening up in the electromagnetic density of states. In contrast to $\lambda/4$ distributed Bragg reflectors, we propose a short-period (subwavelength) metallic/dielectric structure, so that the exciton recombination frequency appears well inside this bandgap. Using metal is essential to enhance the contrast between the refractive indices of layers and suppress coupling to the in-plane guided photon modes.
We assume a spatially-separated cw pump (see, {\it e.g.}, \cite{pump}) at a frequency $\omega_P$ outside the gap which allows optically generated QW excitons to readily relax to lowest-energy states while moving to the central region of the sample.
To obtain the field distribution inside the medium and optimize the layers widths and their number, we numerically solve the Maxwell problem for electric field $\textbf{E}$ inside the structure.

\begin{figure}[t]
\centering
\includegraphics[width=\columnwidth]{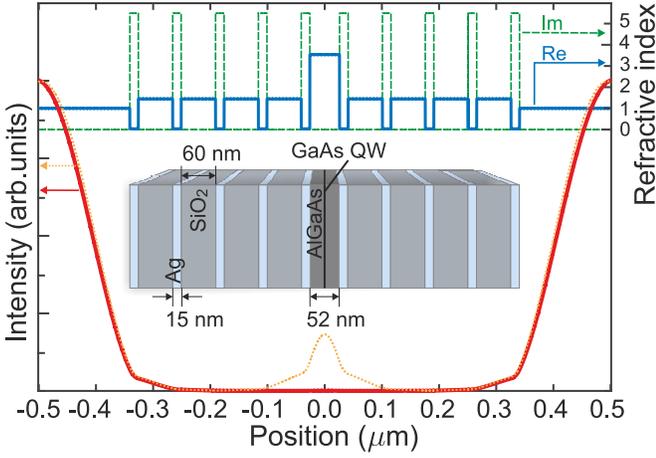}
\caption{Left axis: intensity of field versus $z$, for the schematically illustrated layered structure. Red solid line: for exciton recombination frequency $\omega_X=1.57$~eV$/\hbar$, suppression of light intensity in the region of the QW is $\sim10^3$; orange dotted line: for pump frequency $\omega_P=1.65$~eV$/\hbar$. Right axis: real (blue solid line) and imaginary (green dashed line) parts of the refractive index of the structure $\sqrt{\varepsilon(z)}$ at the frequency $\omega_X$.}
\label{fig01}
\end{figure}

In our work we consider two specific geometries. The first realization is based on a 8-nm GaAs QW embedded in a 52~nm AlGaAs (off-resonant) cavity layer sandwiched between the periodic structure with 4.5 pairs of Ag/SiO$_2$ layers (15/60 nm thickness, respectively), and is depicted schematically in Fig.~\ref{fig01}. 8-nm QWs are well studied in works on IX, so we relate to the data of Ref.~\cite{prb059001625_} with regard to direct exciton recombination line $\hbar\omega_X=1.57$~eV and exciton lifetime without cavity $\tau_{\text{\tiny X}}=70$~ps (recalculated for a single QW). Intensities as a function of $z$, the layers growth direction, are shown in Fig.~\ref{fig01}, for the frequency corresponding to the exciton recombination and the pump frequency ($\hbar\omega_P=1.65$~eV), first being suppressed by a factor of $\sim10^3$, while the latter lies in the region of the cavity resonance (see also Fig.~\ref{fig02}\,(a)) and has a maximum in the region of the QW. The calculation details are given in the Supplemental Material (SM).

Fig.~\ref{fig02} summarizes the optical properties of the proposed structure. Radiative lifetimes for QW excitons inside the cavity were calculated as $\tau_{\text{\tiny X}}$ divided by $\langle|{\bf E}|^2\rangle$ in the region of the QW, for each in-plane wavevector $k$ and frequency $\omega$ in consideration.
Spectral dependence of the emission rate in the normal direction ($k=0$) is shown in Fig.~\ref{fig02}\,(a), and Fig.~\ref{fig02}\,(b) provides inverse lifetimes of excitons with an in-plane wavevector $k$ at the frequency $\omega_X$. The obtained dependence $f(k)=1/\tau(k)$ allows to estimate the radiative lifetime of direct excitons in the system (see below), while for $k=0$ one immediately deduces the lifetime in the ground state: $\tau(0)\approx52$~ns.  Fig.~\ref{fig02}\,(b) also shows that the parasitic optical recombination into the in-plane (guided) photon modes is suppressed ($f(k)\to0$ as $ck/\omega\to1$).
Fig.~\ref{fig02}\,(c) shows the dependence of the lifetime $\tau(0)$ on the number of layers in the structure. The complex dielectric constant of metallic layers results in dissipation of the field, for both $\omega_X$ and $\omega_P$. Thus the optimal number of layers is chosen to provide lifetimes at $\omega_X$ long enough for thermalization, while keeping the line $\omega_P$ still enhanced (the account of losses is given in the SM).

\begin{figure}[t]
\centering
\includegraphics[width=\columnwidth]{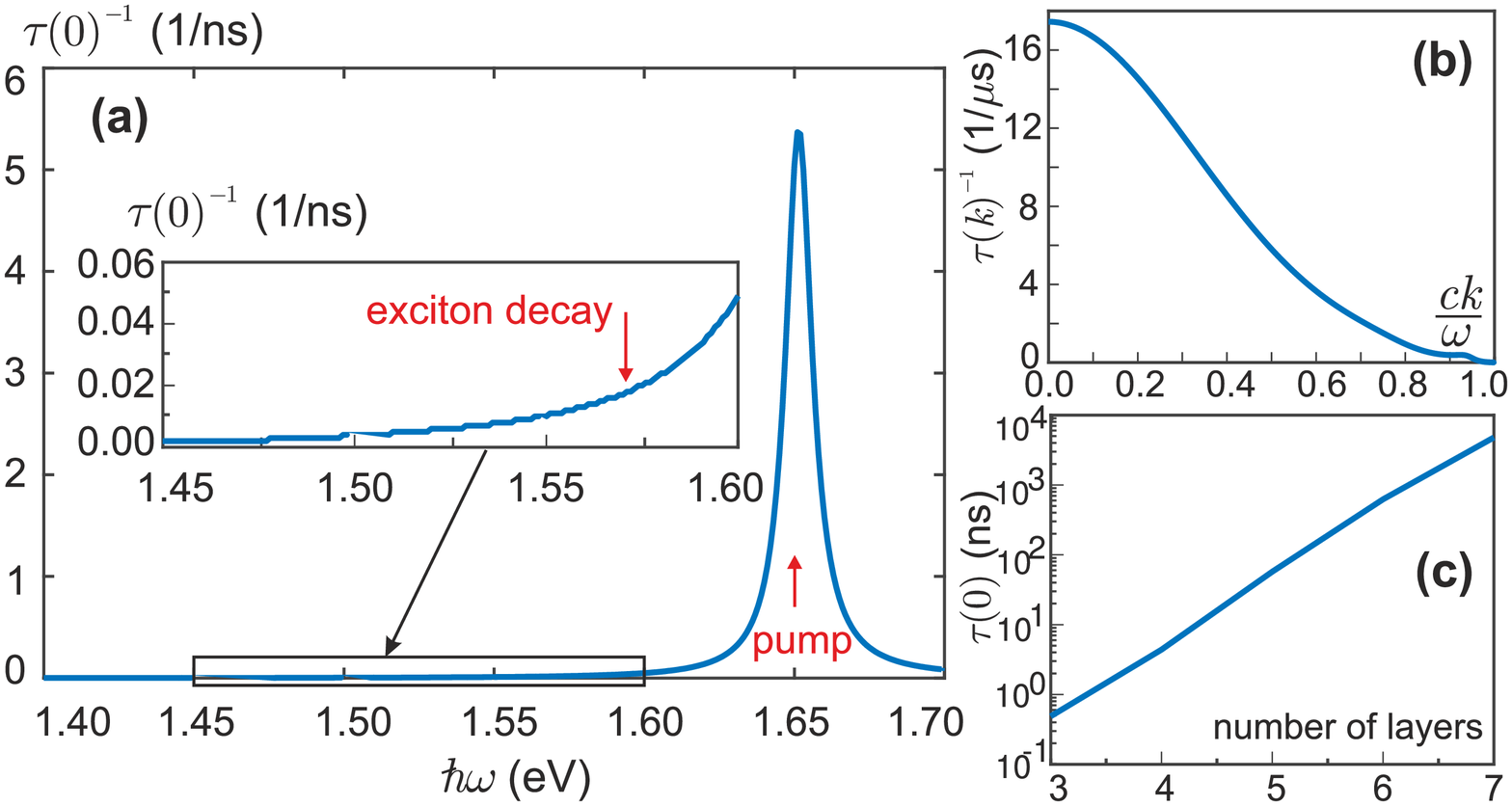}
\caption{Optical properties of the structure shown in Fig.~\ref{fig01}. (a) Spectral dependence of inverse lifetime, deduced from the suppression of emission in the perpendicular direction. Red arrows mark the energies of the pump $\hbar\omega_P=1.65$~eV and exciton optical recombination $\hbar\omega_X=1.57$~eV (inset). For the latter, one deduces the lifetime $\tau(0)\approx52$~ns. (b) Inverse lifetime of excitons with in-plane wavevector $k$ at $\hbar\omega_X$, deduced from angular dependence of emission. Coupling to the in-plane guided modes ($ck/\omega>1$) is absent. (c) Lifetime of ground-state excitons versus number of layers.}
\label{fig02}
\end{figure}

The second geometry we suggest is based on rapidly developing technology allowing to selectively remove substrate and bond thin layers (up to monolayers) of semiconductors. In order to elevate the exciton density and hence the BEC critical temperature, we consider an ultra narrow (2--4 monolayers) single GaAs QW embedded in a 40~nm AlGaAs layer sandwiched between the heterostructure with 4.5 alternate layers of 20/30~nm thick Ag/SiO$_2$. In this ultra-narrow case, fluctuations of the energy bandgap due to fluctuations of Al and Ga concentrations in the AlGaAs barrier lead to a strong disorder. This can be overcome by, \textit{e.g.}, placing on both sides of the QW (i) thin AlAs layers \cite{prb045011403_} or (ii) short-period superlattices $[$(GaAs)$_x$/(AlAs)$_y]_j$ ($x+y=2,3,4$) representing a continuous medium for carriers \cite{prb050014416}.
For this geometry, the recombination energy is estimated as $\hbar\omega_X\simeq1.9$~eV and exciton lifetime without cavity $\tau_{\text{\tiny X}}\sim10$~ps. The lifetime of ground-state excitons in this structure deduced from the Maxwell equations as described above $\tau(0)\approx45$~ps. However, for such a thin QW the effect of dimensionality allows fourfold increase of the exciton density as compared to wider QWs \cite{prb042008928}, so the density can be taken as high as $3.2\cdot10^{11}$~cm$^{-2}$, whereas for the 8-nm QW the lower estimate is $n=8\cdot10^{10}$~cm$^{-2}$ \cite{prb045011403_}.

The considered densities are much higher than IX densities in BEC experiments in CWQs \cite{butov1,butov2_}. At the same time, much smaller exciton Bohr radius ($a_B^X=11$~nm for 8-nm QW and $a_B^X=6$~nm for the ultra-narrow QW) and higher binding energy of direct excitons prevents them from reaching Mott transition (which occurs in CQWs at $n=2\cdot10^{10}$~cm$^{-2}$ and 12--16~K \cite{cqw_mott_tr}).
However, as we confirm below, the exciton gas at those densities is in the regime of intermediate correlations and cannot be readily described by the mean field approximation. To achieve a better analytical description, we unify the Bogoliubov theory with quantum hydrodynamic approach.

\textit{Quantum hydrodynamic description.}---While in macroscopic 2D uniform system BEC is forbidden \cite{hohenberg}, and only superfluid BKT transition takes place \cite{jpc006001181_}, in mesoscopic systems BEC can exist due slow decrease of the density matrix with temperature \cite{kane_}. Therefore we will consider finite but large 2D system of the size $L$, where the disappearance of BEC happens as the BKT crossover \cite{hadzibabic}, and describe the behavior of the equilibrium, one-body density matrix in the long-wavelength (hydrodynamic) limit, \textit{i.e.} at large $r$ (of the order of $L$).
The resulting expression has the form:
\begin{equation}\label{obdm_}
\rho_1({\bf r})\!=\!n\exp\!\left[\frac{1}{S}\!\sum_{{\bf p}\neq0}
\frac{m\varepsilon_{\bf p}\kappa_{\bf p}}{2\tilde n_s p^2}
\frac{e^{\frac{\varepsilon_{\bf p}}{T}}+1}{e^{\frac{\varepsilon_{\bf p}}{T}}-1}
\!\left(\!\cos\frac{\bf p\!\cdot\!r}{\hbar}-1\!\right)\!\right]\!e^{-r/\xi_+},
\end{equation}
where $m$ is the exciton mass, $T$ is their temperature, $S=L^2$ is the area of the quantization, $\tilde{n}_s$ is the superfluid density renormalized by vortex pairs as compared to the uniform superfluid density $n_s$  (we refer to the SM for detailed derivation). Account of free vortices in the system is taken according to Kosterlitz \cite{jpc007001046_} by introducing the factor $e^{-r/\xi_+}$ ($\xi_+$ denotes the distance between free vortices \cite{prl040000783_}).
The Bogoliubov spectrum of excitations $\varepsilon_{\textbf{p}}$ is given by:
\begin{equation}\label{Ep_}
\varepsilon_{\bf p}\equiv
\sqrt{\frac{p^2}{m}\left(\frac{p^2}{4m}+U(\textbf{p})\tilde n_s\right)},
\end{equation}
where $U(\textbf{p})$ contains contributions from two-, three- and many-body interactions in the hydrodynamic Hamiltonian. The constant factor in (\ref{obdm_}) is equal to the total exciton density $n=\rho_1(0)$, and is defined consistently with the UV-cutoff at short distances $\kappa_{\textbf{p}}=(1-p^2/2m\varepsilon_{\textbf{p}})^2$.

In order to estimate the effective interaction $U(\textbf{p})$ in (\ref{Ep_}), one needs to consider a series of ladder diagrams which are dependent on the chemical potential $\mu$ due to the logarithmic divergence of the integrals at the lower limit in the case of small densities ($\mu\rightarrow 0$) \cite{SLY}.
Following Mora and Castin \cite{prl102180404}, we expand the energy functional up to the third order in terms of a small density-dependent parameter $u(n)$, extrapolating the result of \cite{prl102180404} to the crossover regime (\textit{i.e.} intermediate correlations), basing on the comparison of the expansion coefficients with numerical simulations \cite{tbp}.
The bare interaction of direct excitons is described by the Lennard-Jones potential $U_{\rm XX}(r)=W((a^*/r)^{12}-(a^*/r)^6)$, with $a^*\sim a_B^X$. This interaction is short-ranged, so that $U(\textbf{p})$ is weakly dependent on momenta. Hence we assume $U(\textbf{p})\approx U(0)=m^2\chi^{-1}(0)=\partial^2/\partial n^2(E/S)$, $\chi(0)$ being the compressibility of the system. We obtain
\begin{equation}\label{Up_}
U(\textbf{p})\simeq\frac{2\pi\hbar^2}m\frac{d^2(n^2u)}{dn^2},
\end{equation}
where $u=u(n)$ is defined by the transcendental equation $1/u=C_3u-\ln(\pi na_s^2u e^{2\gamma+1/2})$, $a_s>0$ is the 2D wavevector-dependent exciton scattering length \cite{pra081013612_} and $\gamma=0.57721566\dots$ is Euler's constant. The numerical constant $C_3\simeq 2.298\dots$. Solving numerically the equation above for $u$ with the parameters of the suggested structures, one can estimate according to (\ref{Up_}) the dimensionless adiabatic compressibility. For both cases, we get $m^3/4\pi\hbar^2\chi(0)=\partial^2/\partial n^2(n^2u/2)\simeq0.8$ which unambiguously indicates that correlations are not weak.
For better quantitative description of the intermediately-correlated system, we define $a_s$ taking a non-zero wavevector $k\sim1/l_0$, $l_0$ being the healing length.
Note that for a single-component uniform superfluid in the limit of weak correlations ($u\ll1$) and low temperatures ($n-n_s\ll n_s$), expression (\ref{obdm_}) is accurate \cite{tbp}.

The superfluid density $\tilde{n}_s$ in (\ref{obdm_}) is renormalized by the presence of vortex pairs with separations $\lesssim\min(r,\xi_+)$ and can be obtained from the problem of ``dielectric'' screening of the static supercurrent \cite{jpc006001181_}, as follows: $\tilde{n}_s=n_s^l/\epsilon(x_+,a)$. Here $\epsilon(x_+,a)$ is the effective scale-dependent ``dielectric constant'', $a\equiv2\pi\hbar^2n_s^l/mT$, $n_s^l$ is the local superfluid density, and $x_+\equiv\ln[\min(r,\xi_+)/l_0]$ (for details, see the SM). Then, for an infinite 2D system the BKT transition temperature equals $T_c = \pi\hbar^2n_s^l/2m\epsilon_\infty$ \cite{prl039001201_}. However, for a large finite system of the size $L$, the BKT crossover temperature is given by
\begin{equation}\label{TcL_}
T_c^L=\left.\frac{\pi\hbar^2n_s^l(T_c^L)}{2m\epsilon_{\infty}}\right/
\left(1-\frac{\pi^2b^2}{(\ln(L/l_0)+\Delta)^2}\right),
\end{equation}
where the denominator is an analytical fit to the numerical calculation \cite{pla366000487_} with the parameters $\Delta\simeq2.93$, $b\simeq0.80$, and $\epsilon_\infty\simeq1.135$ valid for the 2D $XY$~model.
From (\ref{TcL_}), one obtains the distance between free vortices:
\begin{equation}\label{xi+}
\xi_+\sim\left\{\begin{array}{ll}
\infty,\;&T<T_c,\\
l_0\exp(\pi b/\sqrt{1-T_c/T}-\Delta),\;&T>T_c.
\end{array}\right.
\end{equation}

The local superfluid density $n_s^l$ in Eqs. (\ref{TcL_})--(\ref{xi+}) is calculated with the use of the Landau formula
\begin{equation}\label{nsl_}
n_s^l=n-\sum\limits_{\sigma=1}^{\sigma_{\rm max}}
\int\frac{d\textbf{p}}{(2\pi\hbar)^2}\frac{p^2}{2mT}
\frac{e^{\varepsilon_{\textbf{p}\sigma}/T}}
{(e^{\varepsilon_{\textbf{p}\sigma}/T}-1)^2}
\end{equation}
containing the spectrum of excitations $\varepsilon_{\textbf{p}\sigma}$, where $\sigma$ is the spin index and $\sigma_{\rm max}$ is the spin degeneracy factor.

\begin{figure}[t]
\centering
\includegraphics[width=\columnwidth]{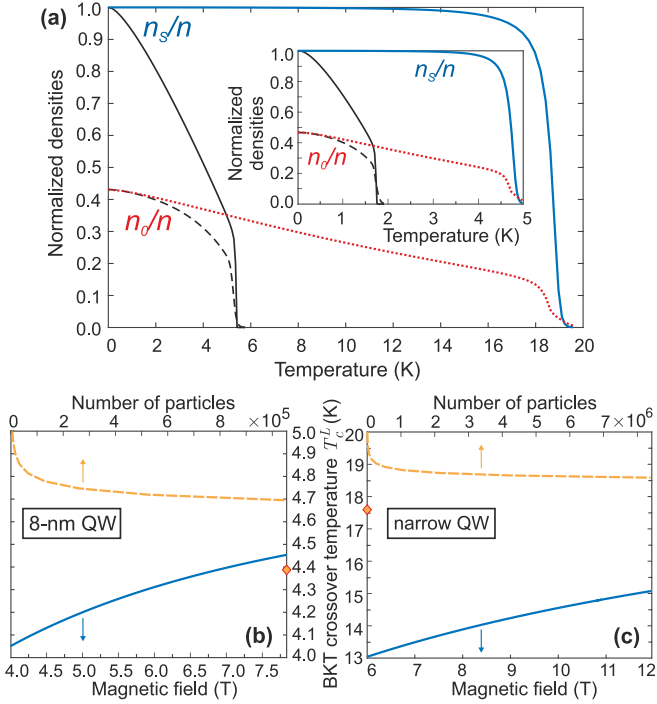}
\caption{(a) Normalized superfluid density $n_s/n$ (blue solid line) and condensate density $n_0/n$ (red dotted line) versus temperature for the ultra-narrow QW, for the case when only one spin component is populated. The BKT crossover is clearly seen at $T_c^L\approx 19$~K. Physical parameters: total number of particles $N=10^6$, $n=3.2\cdot10^{11}$~cm$^{-2}$, $m=0.22m_0$, $a_B^X=6$~nm, $W=10$~meV. Black solid/dashed lines show the same for a multi-component system in the absence of the magnetic field \cite{exchange}.
Inset: same for 8-nm QW, $T_c^L\approx 4.8$~K. Parameters: $N=10^5$, $n=8\cdot10^{10}$~cm$^{-2}$, $m=0.22m_0$, $a_B^X=11$~nm, $W=3$~meV.
(b),\,(c) Top axis, yellow dashed line: critical temperature of the BKT crossover $T_c^L$ versus number of particles in the system (b) for 8-nm QW at $n=8\cdot10^{10}$~cm$^{-2}$ and (c) the ultra-narrow QW at $n=3.2\cdot10^{11}$~cm$^{-2}$. Diamond marker on the vertical axis shows $T_c^L$ for $N=\infty$. Bottom axis, blue solid line: $T_c^L$ versus magnetic field $H$.
}
\label{fig03}
\end{figure}

The results obtained above allow us to evaluate the asymptotic of the one-body density matrix (\ref{obdm_}), the true superfluid density $n_s$, and the condensate density $n_0$ as
\begin{equation}\label{n0_}
n_s=\frac{n_s^l}{\epsilon(\ln(L/l_0),a)}\,,\quad n_0=\frac{1}{S}\int\rho_1(\textbf{r})d\textbf{r}.
\end{equation}
Fig.~\ref{fig03}\,(a) shows the results of calculations for the densities (\ref{n0_}) depending on temperature for the narrow QW realization, revealing $T_c^L\approx19$~K. For 8-nm GaAs QW shown in Fig.~\ref{fig01}, the results are plotted in the inset of Fig.~\ref{fig03}\,(a) with $T_c^L\approx4.8$~K. Dependence of the critical temperature (\ref{TcL_}) on the number of particles at a fixed density is shown in Fig.~\ref{fig03}\,(b) and (c) for 8-nm QW and narrow QW, respectively.

\textit{Accounting for spin.}---So far, excitons were treated as spinless particles with the spectrum $\varepsilon_{\textbf{p}\sigma} \equiv\varepsilon_\textbf{p}$ given by (\ref{Ep_}). Taking into account four spin branches (in GaAs, $\sigma_{\rm max}=4$) with exchange interactions \cite{exchange} lowers $T_c^L$ to 5~K (1.8~K) for the ultra-narrow (8-nm) QW (shown as the black curves in Fig.~\ref{fig03}\,(a)). The spinless approximation however can be justified by employing the Zeeman effect.
In order to analyze quantitatively at which magnetic fields one can neglect spin, we solve the BKT transition problem in magnetic field $H$ to evaluate the spectrum $\varepsilon_{\textbf{p}\sigma}$ in dilute and low-temperature limit. In this case, the lowest branch is given by $\varepsilon_{\textbf{p}1}=\varepsilon_{\textbf{p}}$, while the higher branches have the form $\varepsilon_{\textbf{p}\sigma}= p^2/2m+D_{\sigma}$. Here $D_{\sigma}>0$ are the Zeeman shifts in which all $g$--factors are taken to be equal 1, so that $D_{\sigma}/(e\hbar H/2mc)=1,3,4$ at $\sigma=2,3,4$.
As one can see from the simulation results shown in Fig.~\ref{fig03}\,(b) for the 8-nm QW, even for moderate fields the depletion of the superfluid component is low: $T_c^L$ is lowered less than by $15\%$ at $H=4$~T, and less than by $10\%$ at $H=6$~T. This underlines the consistency of our spinless approximation.
In the ultra-narrow QW, the depletion of the superfluid component by magnetic field is more pronounced but still moderate (see Fig.~\ref{fig03}\,(c)).

Finally, the exciton lifetime is defined by
\begin{equation}\label{lum}
\frac1{\tau}=\frac1{\tau(0)}\int\frac{d\textbf{r}d\textbf{k}}{(2\pi)^2n} \, \rho_1(\textbf{r}) e^{-r/\xi}\frac{f(k)}{f(0)}e^{i{\bf k\cdot r}},
\end{equation}
where $f(k)$ is given in Fig.~\ref{fig02}\,(b). The factor $e^{-r/\xi}$ indicates that the system is not fully thermalized at large scales: in thermal equilibrium $\xi\to\infty$.
According to (\ref{lum}), for the 8-nm (ultra-narrow) QW we get $\tau\approx150$ (140)~ns. Keeping in mind that in CQWs, within the IX lifetimes $\tau_{\text{\tiny IX}}\sim100$~ns \cite{prb085045207}, BEC occurs on the scales of the order of $12\,\mu$m \cite{butov2_}, one concludes that for our structures, in the system of the size $L\sim\sqrt{N/n}$, the achieved lifetime is {\it a fortiori} long enough for reaching condensation.

It is important to note that finite lifetime does not affect the superfluidity in the system and the employed hydrodynamic formalism. Indeed, the time required for a wavepacket to pass with the sound velocity  $c_s=\sqrt{mn_s/\chi(0)}$ from one side of the system to the other $t\sim L/c_s$ is $10^3$ times shorter than $\tau$. Hence the sound damping is negligible, while the flow velocity $v\sim L/\tau$ produced by the exciton decay is 3 orders of magnitude less than Landau critical velocity $v_c\sim c_s$.

\begin{figure}[b]
\centering
\includegraphics[width=\columnwidth]{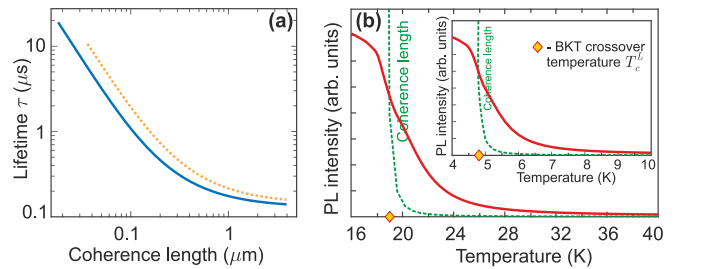}
\caption{(a) Lifetime of excitons with small momenta versus coherence length (blue solid line: for ultra-narrow QW, yellow dotted line: for 8-nm QW). (b) Intensity of PL in direction close to normal (red solid line) versus temperature, for ultra-narrow QW in thermal equilibrium. The increase of intensity below $T_c^L$ is 82 times. Green dashed line shows the coherence length versus $T$ in arbitrary units, revealing its rise to infinity at $T_c^L$. Inset: same for 8-nm QW. The increase of intensity below $T_c^L$ is 45 times. Parameters are the same as in Fig.~\ref{fig03}.}
\label{fig04}
\end{figure}
We complete our analysis by estimating the change in PL intensity in the region of the BKT crossover.
During thermalization, $\tau$ drops after the appearance of the quasi-condensed phase. In particular, as one can see in Fig.~\ref{fig04}\,(a), in the absence of quasi-condensate ($\xi_+\sim1/\sqrt n$), $\tau$ is approximately two orders of magnitude longer than when coherence length is $\sim1\mu$m. The reason for this effect is the drastic narrowing of the $k$-distribution of the system with the growth of the coherence length and appearance of quasi-condensate, which pushes all excitons into the cavity radiative region. As a result, excitons start to actively recombine which can be seen in PL.
Fig.~\ref{fig04}\,(b) displays sharp increase of the coherence length in thermalized system at the crossover and the corresponding manyfold growth of the emission intensity from the structure: as compared to intensity at $T>T_c^L$, it is 45 and 82 times higher below $T_c^L$ for 8-nm QW and ultra-narrow QW, respectively.

It should be mentioned that we assume the QWs to be of a high quality with low inhomogeneities, {\it e.g.} same as used in Ref. \cite{butov2_}. However, when disorder is taken into account, critical temperatures are still estimated to be 4~K and higher (see the SM).

In conclusion, we proposed the method to increase the lifetime of direct excitons in single GaAs QWs by employing photonic engineering, and predict their transition to the superfluid phase and condensation at temperatures from 4.8~K to 19~K, depending on the geometry. For comparison, these values are well above $T_c=0.1$~K demonstrated in CQWs \cite{butov2_}, and are not lower than the temperatures theoretically estimated for IX superfluidity in spatially-separated MoS$_2$ layers \cite{butov3}. We would like to note that for a transition metal dichalcogenide (TMD) monolayer embedded in an off-resonant cavity, our theory predicts $T_c^L$ as high as 85~K. However, while the optical recombination of excitons would be suppressed, Auger processes being dominant in TMDs \cite{glazov} present the main obstacle for exciton relaxation.

\begin{acknowledgments}
\textit{Acknowledgements.}---The authors are grateful to Oleg Kotov for discussions. N.S.V. acknowledges the financial support by the Russian Foundation for Basic Research, according to the research Project No. 16--32--60066 mol\_a\_dk, and the Council of the President of the Russian Federation for Support of Young Scientists and Scientific Schools (Project No. MK--201.2017.2). Yu.E.L. is supported by the Program of Basic Research of the High School of Economics.
\end{acknowledgments}

\clearpage

%\onecolumn
\begin{widetext}
\begin{centering}
\large{\bf Bose condensation of direct excitons in an off-resonant cavity at elevated temperatures --- Supplemental Material}
%\vspace{10pt}
\begin{quote}
{\small In this supplemental material, we provide additional figures and details of theoretical calculations, and address the effects of disorder.}
\end{quote}
\end{centering}

\section{Maxwell description and Ultra narrow quantum well}

In order to obtain field distribution inside the structure, we solve the Maxwell problem $(\Delta - \varepsilon/c^2\,\partial_{tt})\textbf{E}=0$ for the 3D vector of electric field $\textbf{E}$, using the Runge--Kutta method. Here $\Delta$ denotes the three-dimensional vector Laplacian, $\varepsilon$ is the medium dielectric constant, and $c$ is the velocity of light in vacuum.
Given the in-plane translational symmetry of the structure (shown in Fig.~1 of the main text for 8-nm QW and in Fig.~\ref{sup_fig01}(b) of this Supplemental material for the ultra-narrow QW), we obtain $\textbf{E}=e^{i\omega t}e^{i\textbf{k}\cdot\textbf{r}}\textbf{E}(z)$, where $z$ is the layers growth direction, $\omega$ is the frequency of the field, and $\textbf{k}$ and $\textbf{r}$ are the 2D in-plane wavevector and radius-vector, respectively. As a result, we are solving the equation
\begin{equation}
\left(\frac{\partial^2}{\partial z^2}-k^2+\frac{\omega^2}{c^2}\,\varepsilon(z)\right)\textbf{E}(z)=0.
\end{equation}
with the complex dielectric constant $\varepsilon(z)$ of the structure corresponding to the given frequency. In particular, for silver layers we used the following data: $\sqrt{\varepsilon} = 0.0351659+i\cdot5.49339$ at $\hbar\omega=1.57$~eV (as shown in Fig.~1 of the main text); $\sqrt{\varepsilon} = 0.031108+i\cdot5.20538$ at $\hbar\omega=1.65$~eV; $\sqrt{\varepsilon} = 0.0515819+i\cdot4.42928$ at $\hbar\omega=1.9$~eV (as shown in Fig.~\ref{sup_fig01}(a)); and $\sqrt{\varepsilon} = 0.0540027+i\cdot3.9762$ at $\hbar\omega=2.08$~eV.
%Since the frequency of the exciton decay (pump) $\omega_{X(P)}$ is fixed, profile of the magnetic vector potential ${\bf A}=(-\sqrt{\varepsilon}/c)\partial{\bf E}/\partial t$ up to $\sqrt{\varepsilon}$ coincides with the profile of ${\bf E}$. Therefore
$\langle|{\bf E}|^2(z)\rangle$ was normalized to unity (here
$\langle f(z)\rangle=\int_0^{\lambda}f(z)dz/\lambda$ denotes averaging over the period $\lambda=2\pi c/\omega$).% over the medium with excitons rather than in free space, {\it i.e.} in GaAs with $\varepsilon=12.5$.
\begin{figure}[b]
\centering
\includegraphics[width=\textwidth]{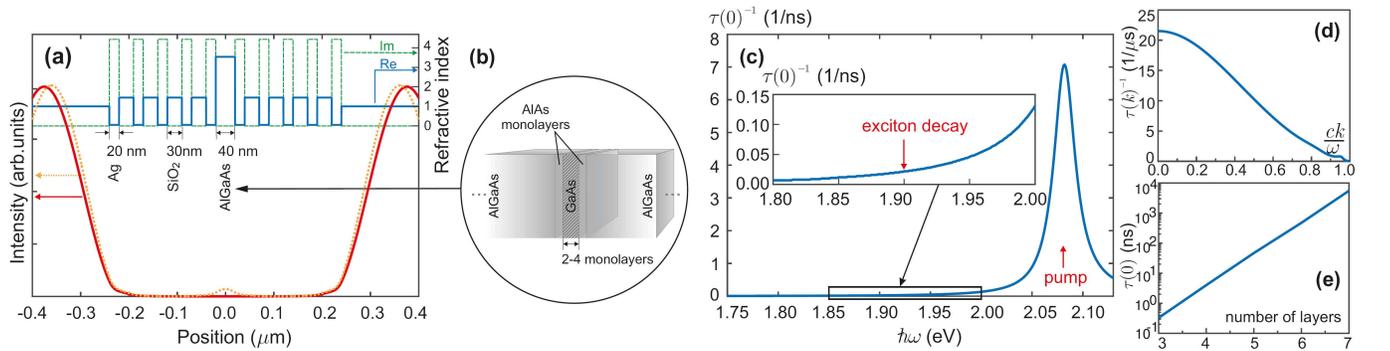}
\caption{(a) Left axis: intensity of field versus $z$, for the ultra-narrow QW (see details in main text). Red solid line: for exciton recombination frequency $\omega_X=1.9$~eV$/\hbar$, suppression of light intensity in the region of the QW is $4.4\times10^3$; orange dotted line: for pump frequency $\omega_P=2.08$~eV$/\hbar$. Right axis: real (blue solid line) and imaginary (green dashed line) parts of the refractive index of the structure at the frequency $\omega_X$. (b) Schematic illustration of the central layer of the structure $\sqrt{\varepsilon(z)}$ with QW placed between the two AlAs monolayers, in order to reduce fluctuations of the energy bandgap due to fluctuations of Al and Ga concentrations in the AlGaAs barrier. (c) Spectral dependence of inverse lifetime. Red arrows mark the frequencies of the pump $\hbar\omega_P=2.08$~eV and exciton optical recombination $\hbar\omega_X=1.9$~eV (shown in the inset). For $1.9$~eV, the deduced lifetime $\tau(0)\approx45$~ns. (d) Inverse lifetime of excitons with in-plane wavevector $k$ at $\hbar\omega_X$. (e) Lifetime of ground-state excitons versus number of layers.}
\label{sup_fig01}
\end{figure}
Comparing the resulting profiles of the electric field intensity for different $\textbf{k}$ (corresponding to the angle of emission) to $\langle|{\bf E}|^2\rangle$ in vacuum, we obtain suppression of the field.
Given the lifetimes of the QW exciton without cavity (70~ps for 8-nm QW \cite{prb059001625} and 10~ps for ultra-narrow QW (estimate)) and the suppression of field intensity, we calculate the lifetimes of excitons with different $k$ inside the structure, dividing the lifetime without cavity by $\langle|{\bf E}|^2\rangle$ in the region of the QW. Results of the calculations for the 8-nm QW are shown in the main text. For the ultra-narrow QW, the corresponding results are given in Fig.~\ref{sup_fig01}~(a), (c), (d), and (e). Fig.~\ref{sup_fig01}(b) schematically shows one of the suggested ways to overcome disorder due to fluctuations of Al and Ga atoms concentrations in the AlGaAs barrier, by placing AlAs monolayers by both sides of the ultra-narrow QW \cite{prb045011403}. The disorder caused by the fluctuations of QW widths is suppressed when the AlGaAs (or GaAs) heterointerface is capped by exact number of monolayers.

The reason to use metallic layers in the proposed structures was to eliminate the guided (in-plane) photon modes which inevitably appeared when the structures were composed of dielectric layers. However, metallic layers have a complex dielectric constant as given above, and therefore cause dissipation of the field. To estimate the effect of losses, we have performed the same calculation as presented in Fig.~2(a) of the main text and Fig.~\ref{sup_fig01}(c) of this Supplemental Material, varying the number of alternate layers in the structures and hence the amount of losses. Results in Fig.~\ref{sup_fig03} display the spectral dependence of inverse lifetime for 8-nm QW, {\it i.e.} the same curve as in Fig.~2(a) of the main text, for the number of metallic layers $N=1$, $2$, $3$, $5$, $7$, $10$ by each side of the cavity (AlGaAs) layer, both in normal scale (a) for comparison and in semilogarithmic scale (b) for visibility.
\begin{figure}[h]
\centering
\includegraphics[width=0.65\textwidth]{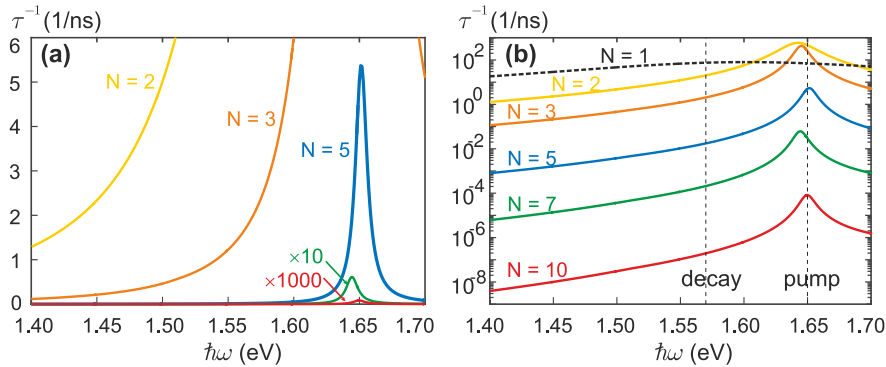}
\caption{(a),(b) Spectral dependence of inverse lifetime, deduced from the suppression of emission in the perpendicular direction from the structure with embedded 8-nm QW as described in the main text. Black dashed line: 1 Ag layer by the each side of the cavity; yellow solid line: 2 Ag layers; orange: 3 Ag layers; blue: 5 Ag layers (same curve as shown in Fig.~2(a) of the main text); green: 7 Ag layers; red: 10 Ag layers.}
\label{sup_fig03}
\end{figure}
As one can clearly see, starting from $N=2$, the bandgap occurs and the cavity layer starts enhancing the pump frequency $\omega_P=1.65$~eV$/\hbar$, those effects being more and more pronounced with the increase of the number of layers. Starting from $N=3$, the intensity of field at the exciton recombination frequency $\omega_X=1.57$~eV$/\hbar$ is several orders of magnitude lower than at $\omega_P$. With the growth of $N$, due to losses in metal, at all frequencies the intensity decreases, and at $N=7$, $10$, one sees that the pump intensity at $\hbar\omega_P$, while still displaying a peak due to the cavity enhancement, appears also extremely low. This interplay defines the choice of the optimal number of layers: (i) the lifetime of excitons at $\omega_X$ should be long enough for thermalization (but not too long to avoid non-radiative recombination which happens on $\mu$s-timescales \cite{kukushkin1}), and (ii) the intensity of field at $\omega_P$ should be high enough to allow effective pumping and hence populating the system with excitons.

%%%%%%%%%%%%%%%%%%%%%%%%%%%%%%%%%%%%%%%%%%%%%%%%%%%%%%%%%%%%%%%%%%%%%%%%%%%%%%
\section{Quantum hydrodynamic description with Bogoliubov spectrum}

Describing coherence in the system, we are interested in the behavior of the equilibrium, one-body density matrix $\rho_1(\textbf{r})\equiv\langle\hat\Psi^\dag(\textbf{r})\hat\Psi(0)\rangle$ in the long-wavelength (hydrodynamic) limit, \textit{i.e.} at large $r\sim L$.
Here $\hat\Psi(\textbf{r})$ is the exciton field operator and the brackets $\langle\cdot\rangle$ denote the averaging over the thermal equilibrium state of a 2D uniform exciton system with the constant number of particles. To find the asymptotic behavior of $\rho_1(\textbf{r})$, we substitute the field operator as
\begin{equation}\label{psis}
\hat{\Psi}(\textbf{r})=\exp(i\hat\varphi(\textbf{r}))\sqrt{\hat{\rho}(\textbf{r})}, \, \quad \hat{\Psi}^\dag(\textbf{r})=\sqrt{\hat{\rho}(\textbf{r})}\exp(-i\hat\varphi(\textbf{r})),
\end{equation}
where $\hat{\rho}(\textbf{r})=n_s+\hat\rho^\prime(\textbf{r})$ and $\hat\varphi(\textbf{r})$ are the superfluid density and phase operators, respectively, $n_s$ being the uniform superfluid density and $\hat\rho^\prime(\textbf{r})$ the density fluctuations operator.

Kubo cumulant expansion \cite{kubo,kane} up to the second term provides for the average
\begin{equation}\label{cumulant}
\langle \exp\left[-i(\hat\varphi(\textbf{r}) - \hat\varphi(0))\right]\rangle = \exp\left[-\frac{1}{2}\langle(\hat\varphi({\bf r})-\hat\varphi(0))^2\rangle\right].
\end{equation}

To avoid the explicit use of the phase operator which is not well defined \cite{rmp040000411}, in our calculations we will instead work with the
operator of superfluid velocity $\hat{\textbf{v}}(\textbf{r})=\hbar\nabla\hat\varphi(\textbf{r})/m$. Splitting it into longitudinal (phonon) and transverse (vortical) parts as $\hat{\bf v}({\bf r})=\hat{\bf v}_{\parallel}({\bf r})+\hat{\bf v}_{\perp}({\bf r})$, we assume $\hat{\bf v}_{\perp}({\bf r})=0$ and treat vortical effects by introducing appropriate renormalization. In particular, we take account of free vortices in the system (if those are present)
in the form of Kosterlitz renormalization \cite{jpc007001046}, while the renormalization by vortex pairs can be accounted for inside $n_s$ ($n_s\,\rightarrow\,\tilde{n}_s$) \cite{prb024002526}. The resulting expression that allows to explicitly evaluate $\rho_1({\bf r})$ in the long-wavelength limit is as follows:
\begin{equation}\label{obdm2}
\rho_1({\bf r})={\rm const}\cdot \text{exp}\!\left[\textstyle-\!\left\langle\!\left(\frac m{\hbar}\int\limits_0^{\bf r}(\hat{\bf v}_{\parallel}({\bf r}')d{\bf r}')\right)^2\!\!/2 \right\rangle\!\right]\!e^{-r/\xi_+}.
\end{equation}
Here $\xi_+$ denotes the distance between free vortices \cite{prl040000783} (in the case of no free vortices $\xi_+=\infty$), and the constant factor in front of the exponent depends on the choice of the ultraviolet (UV) cutoff and will be defined below. % at $\textbf{p}\rightarrow 0$ ($\textbf{r}\rightarrow\infty$).
Note that the integral in (\ref{obdm2}) does not depend on the choice of the contour connecting the points $0$ and $\textbf{r}$ \cite{prb024002526}.

Making use of the Fourier transform, one has
\begin{equation}\label{phi0r}
\frac m{\hbar}\int\limits_0^{\bf r}(\hat{\bf v}_{\parallel}({\bf r}')d{\bf r}') = \sum_{{\bf p}\ne0}\frac{m{\bf p}}{ip^2}\hat{\bf v}_{\bf p}
\frac{e^{i{\bf p\cdot r}/\hbar}{-}1}{\sqrt S},
\end{equation}
where $\hat{\bf v}_{\bf p}=
\int e^{-i{\bf p\cdot r}/\hbar}\hat{\bf v}_{\parallel}({\bf r})d{\bf r}/\sqrt S$, $S$ is the area of quantization, and we took into account $\hat{\bf v}_{\bf p}\parallel{\bf p}$.

To calculate the average in (\ref{obdm2}), we substitute (\ref{psis}) into the grand canonical Hamiltonian
\begin{equation}\label{hamiltonian}
\hat H -\mu\hat N = \int\hat\Psi^\dag(\textbf{r})\left(-\frac{\hbar^2}{2m}\Delta - \mu\right)\hat\Psi(\textbf{r})d\textbf{r} +  \frac{1}{2}\iint\hat\Psi^\dag(\textbf{r})\hat\Psi^\dag(\textbf{r}^\prime) U_2(\textbf{r}-\textbf{r}^\prime)\hat\Psi(\textbf{r}^\prime)\hat\Psi(\textbf{r}) d\textbf{r}^\prime d\textbf{r} + \hat{U}_3 + \dots,
\end{equation}
where $m$ is the exciton mass, $\mu$ is their chemical potential, $U_2(\textbf{r})$ is the exciton-exciton two-particle interaction, and $\hat U_3 + \dots$ denote three-body and other many-body interaction operators.
After transformations, one can separate the harmonic part of the hydrodynamic Hamiltonian (\ref{hamiltonian}) as $\hat H-\mu\hat N = \text{const} + \hat{H}_0 + \hat V$ (see \cite{prb049001205,prb049012938}),
where $\hat V$ contains all anharmonic terms and
\begin{equation}\label{h0}
\hat{H}_0 = \sum\limits_{\textbf{p}\neq0} \left(\frac{m\tilde{n}_s}{2}\,\hat{\textbf{v}}_{\textbf{p}}\cdot \hat{\textbf{v}}_{-\textbf{p}} + \left(\frac{p^2}{8m\tilde{n}_s}+\frac{U(\textbf{p})}{2}\right)\hat{\rho}_{\textbf{p}} \hat{\rho}_{-\textbf{p}}\!\right).
\end{equation}
In (\ref{h0}), $\hat{\rho}_{\textbf{p}}$ denotes the Fourier transforms of the density fluctuations operator $\hat{\rho}^\prime(\textbf{r})$, and $U(\textbf{p})$ contains contributions from two-, three- and many-body interactions in the hydrodynamic Hamiltonian (\ref{hamiltonian}).
The transformation
\begin{equation}\label{vp}
\hat{\rho}_{\textbf{p}}=i\sqrt{\frac{\tilde{n}_s p^2}{2m\varepsilon_{\bf p}}}\left(\hat c_{\textbf{p}}-\hat{c}^\dag_{-\textbf{p}}\!\right)\!,\, \quad
\hat{\textbf{v}}_{\textbf{p}}=\frac{i\textbf{p}}{m}\sqrt{\frac{\varepsilon_{\bf p}m}{2\tilde n_s p^2}} \left(\hat c_{\textbf{p}} + \hat{c}^\dag_{-\textbf{p}}\!\right)
\end{equation}
brings (\ref{h0}) to the diagonalized form $\hat H_0=\sum\limits_{{\bf p}\ne0}\varepsilon_{\bf p} \hat c_{\bf p}^\dag\hat c_{\bf p}$, where $\hat c_{\textbf{p}}$ is the annihilation operator of a collective phonon with the momentum $\textbf{p}$ satisfying bosonic commutation relations $[\hat c_{\textbf{p}},\hat c_{\textbf{p}^\prime}]=0$, $[\hat c_{\textbf{p}},\hat c_{\textbf{p}^\prime}^\dag]=\delta_{\textbf{pp}^\prime}$, and $\varepsilon_{\textbf{p}}$ is the Bogoliubov spectrum of excitations:
\begin{equation}\label{Ep}
\varepsilon_{\bf p}\equiv
\sqrt{\frac{p^2}{m}\left(\frac{p^2}{4m}+U({\bf p})\tilde n_s\right)}.
\end{equation}

Substituting (\ref{vp}) into (\ref{phi0r}) and (\ref{obdm2}), and taking into account $\langle\hat c_{\textbf{p}}\rangle=\langle\hat c_{\textbf{p}}\hat c_{\textbf{p}^\prime}\rangle=0$ and $\langle\hat c_{\textbf{p}}^\dag\hat c_{\textbf{p}^\prime}
\rangle=\delta_{\textbf{pp}^\prime}/(e^{\varepsilon_{\textbf{p}}/T}-1)$, where $T$ is the temperature of excitons, one gets the result
\begin{equation}\label{rho1}
\rho_1(\textbf{r})\! =\!\text{const}\cdot\exp\!\left[\sum_{{\bf p}\neq0}\!
\frac{m\varepsilon_{\bf p}}{2\tilde n_s p^2S}
\frac{e^{\frac{\varepsilon_{\bf p}}{T}}+1}{e^{\frac{\varepsilon_{\bf p}}{T}}-1}
\!\left(\!\cos\frac{\bf p\cdot r}{\hbar}{-}1\right)\!\right]\!e^{-r/\xi_+}.
\end{equation}
It should be pointed out that the hydrodynamic quantum field theory approach described above is valid only in the long-wavelength limit $p\rightarrow 0$. In other words, the accuracy of the result (\ref{rho1})
is limited by the choice of the cutoff at short distances. Hence our approach involves two phenomenological parameters, the constant factor that is contained in (\ref{obdm2}) and the UV-cutoff factor $\kappa_{\textbf{p}}$:
\begin{equation}\label{kappa}
\kappa_{\textbf{p}}(\textbf{r})=\left\{
\begin{array}{lcl}
1,\,p\rightarrow 0,\\
0,\,p\rightarrow\infty
\end{array}
\right.
\end{equation}
that cannot be calculated analytically within the hydrodynamic description \cite{Popov} and has to be introduced under the sum in (\ref{rho1})
by hand. These two parameters are to be defined consistently with each other.
Choosing $\kappa_{\textbf{p}}(\textbf{r})$ in the form
\begin{equation}\label{kp}
\kappa_{\textbf{p}}=\left(1-\frac{p^2}{2m\varepsilon_{\textbf{p}}}\right)^2
\end{equation}
leads to the constant in (\ref{obdm2}) being equal to the full exciton density $n=\rho_1(0)$. Thus we obtain the long-wavelength asymptotic for the one-body density matrix which appears in the main text:
\begin{equation}\label{obdm}
\rho_1({\bf r})\!=\!n\exp\!\left[\frac{1}{S}\!\sum_{{\bf p}\neq0}
\frac{m\varepsilon_{\bf p}\kappa_{\bf p}}{2\tilde n_s p^2}
\frac{e^{\frac{\varepsilon_{\bf p}}{T}}+1}{e^{\frac{\varepsilon_{\bf p}}{T}}-1}
\!\left(\cos\frac{\bf p\!\cdot\!r}{\hbar}-1\right)\!\right]\!e^{-r/\xi_+}.
\end{equation}

%%%%%%%%%%%%%%%%%%%%%%%%%%%%%%%%%%%%%%%%%%%%%%%%%%%%%%%%%%%%%%%%%%%%%%%%
\section{Dielectric screening and superfluid density renormalization}

As the renormalization due to free vortices is taken into account according to Kosterlitz \cite{jpc007001046} by introducing the factor $e^{-r/\xi_+}$ (see (\ref{obdm2}), (\ref{obdm})), the superfluid density $\tilde{n}_s$ renormalized only by the presence of vortex pairs and is obtained from the problem of ``dielectric'' screening of the static supercurrent \cite{jpc006001181}. The effective scale-dependent ``dielectric constant'' of this screening for 2D $XY$~model satisfies the equation
\begin{equation}\label{e(x,a)}
\frac{\partial\epsilon(x,a)}{\partial x}=\pi^2ae^{-\pi a/2}\exp\!\left(\!
4x-a\!\int\limits_0^x\!\frac{dx^\prime}{\epsilon(x^\prime,a)}\!\right).
\end{equation}
In (\ref{e(x,a)}), $a\equiv2\pi\hbar^2n_s^l/mT$, $n_s^l$ is the local superfluid density, and $x=\ln(l/l_0)$, $l$ denoting the separation between vortices and $l_0$ the microscopic phonon length scale of the order of the vortex core radius, {\it i.e.} the healing length (for a realistic system of 2D excitons, $l_0\sim 1/\sqrt{n}$). When $l=l_0$ (\textit{i.e.} $x=0$), the interaction of a pair of vortices separated by $l_0$ is unaffected by any other pair, which provides the boundary condition for (\ref{e(x,a)}): $\epsilon(0,a)=1$.

For one-body density matrix between the points $0$ and $\textbf{r}$, all pairs with $l>r$ do not take part in the renormalization of the superfluid density \cite{prb024002526}. On the other hand, if $l$ is larger than the distance between free vortices $\xi_+$, the vortex pair itself should be considered as two free vortices. Hence the global superfluid density $\tilde n_s$ in (\ref{obdm}) is renormalized only by the pairs with separation $l\lesssim\min(r,\xi_+)$ and depends on $r$ as
\begin{equation}\label{nsp}
\tilde{n}_s=\frac{n_s^l}{\epsilon(x_+,a)},\quad x_+\equiv\ln\frac{\min(r,\xi_+)}{l_0}\,,
\end{equation}
while the true superfluid density $n_s$ which is correspondent to the screening on the scales of the order of the system size $L$, is given by
\begin{equation}\label{ns}
n_s=\frac{n_s^l}{\epsilon(\ln(L/l_0),a)}\,.
\end{equation}

For an infinite system the BKT transition temperature is given by $T_c = \pi\hbar^2n_s^l/2m\epsilon_\infty$ \cite{prl039001201} (where $\epsilon_\infty = \epsilon(\infty, 4\epsilon_\infty)$). However, for a large finite system of the size $L$ the BKT crossover temperature $T_c^L$ is obtained from the numerical analysis of Eqs. (\ref{e(x,a)}), (\ref{ns}) in the assumption $\ln(L/l_0)\gg1$ \cite{pla366000487}:
\begin{equation}\label{TcL}
T_c^L=\left.\frac{\pi\hbar^2n_s^l(T_c^L)}{2m\epsilon_{\infty}}\right/
\left(1-\frac{\pi^2b^2}{(\ln(L/l_0)+\Delta)^2}\right).
\end{equation}
The denominator in (\ref{TcL}) is an analytical fit to the numerical calculation \cite{pla366000487} with the parameters $\Delta\simeq2.93$, $b\simeq0.80$, and $\epsilon_\infty\simeq1.135$ valid for the 2D $XY$~model. Since the BKT crossover corresponds to the appearance of free vortices in the system, for the system of the size $L$ it occurs when $\xi_+\sim L$. This allows to obtains the distance between free vortices $\xi_+$ from (\ref{TcL}) (see Eq. (5) of the main text).

%%%%%%%%%%%%%%%%%%%%%%%%%%%%%%%%%%%%%%%%%%%%%%%%%%%%%%%%%%%%%%%%%%%%%%%%
\newpage
\section{Disorder}

\begin{wrapfigure}[21]{r}{200pt}
\centering
\includegraphics[width=0.35\textwidth]{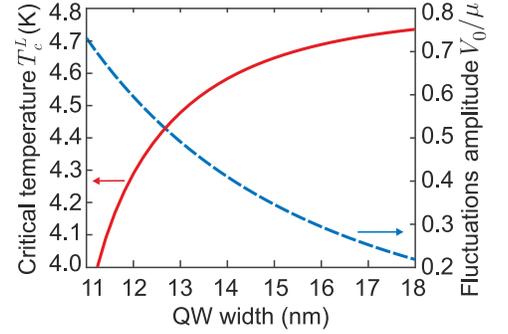}
\caption{Disorder effects for medium quality quantum wells. Left vertical axes, red solid line: BKT crossover temperature $T_c^L$ versus cavity width $L_{\rm QW}$. For wide QWs the depletion of the superfluid component by disorder essentially saturated and the critical temperature tends to its calculated value 4.8~K. Right vertical axis, blue dashed line: normalized amplitude of random field fluctuations for the QW width fluctuations of one monolayer.}
\label{sup_fig02}
\end{wrapfigure}
While in this work we assume QWs to be of a high quality with low inhomogeneities, {\it e.g.} same as reported in Refs. \cite{prb059001625,butov2}, for completeness of our analysis we estimate the effects of disorder. In order to do so, we take local superfluid density $n_s^l$ as given by Eq.~(6) of the main text and substract from  $n_s^l$ the quantity
\begin{equation}
\frac{V_0^2\Lambda^2}{2m^2\hbar^2}\int\limits_0^{\infty}pdp\,
\frac{p^4}{\varepsilon_p^4}e^{p^2\Lambda^2/2\hbar^2}
\end{equation}
that represents the depletion of the superfluid component by disorder  \cite{prb049012938}. Here $\lambda$ and $V_0$ are the width and the amplitude of the random field fluctuations, respectively:
$$\langle\langle V(\textbf{r})V(\textbf{s})\rangle\rangle=V_0^2e^{-(\textbf{r}-\textbf{s})^2/2 \Lambda^2}.$$
For medium quality of the QW, we take $\Lambda=0.5$~$\mu$m and assume the amplitude $V_0=\pi^2\hbar^2d_0/(m'(L_{\rm QW}^{\rm eff})^3)$ to correspond to the fluctuations of the QW width $L_{\rm QW}$ by one monolayer $d_0=0.283$~nm (where $m'=m_em_h/m=0.046m_0$ is the reduced mass of an electron and a hole, and the effective QW width $L_{\rm QW}^{\rm eff}=L_{\rm QW}+2\Delta_t$ takes into account the tunneling of carriers under the AlGaAs barrier, $\Delta_t=\hbar/\sqrt{2m_{e,h}U_{e,h}}\approx1.5$~nm). Fig.~\ref{sup_fig02} shows the calculations result for the QWs of the width in the range from $L_{\rm QW}$=11~nm to 18~nm. For wider QWs, the effect of disorder is negligible. For more narrow QWs such as considered in our manuscript, the high quality of the structures is essential. However even when disorder is taken into account, the critical temperature is estimated to be 4~K and higher.

\end{widetext}


\begin{thebibliography}{99}
\bibitem{keldysh}
L. V. Keldysh, A. N. Kozlov, Sov. Phys. JETP \textbf{27}, 521--528 (1968).

\bibitem{lozovik}
Yu. E. Lozovik, V. I. Yudson, Sov. Phys. JETP \textbf{44}, 389--397 (1976).

\bibitem{rice}
X. Zhu, P. B. Littlewood, M. S. Hybertsen, and T. M. Rice, Phys. Rev. Lett. \textbf{74}, 1633 (1995).

\bibitem{exc2}
J. L. Lin and J. P. Wolfe, Phys. Rev. Lett. \textbf{71}, 1222 (1993).

\bibitem{exc1}
T. Fukuzawa, E. E. Mendez, and J. M. Hong, Phys. Rev. Lett. \textbf{64}, 3066 (1990).

\bibitem{exc3}
L. V. Butov and A. I. Filin, Phys. Rev. B \textbf{58}, 1980 (1998).

\bibitem{butov1}
L. V. Butov, C. W. Lai, A. L. Ivanov, A. C. Gossard, and D. S. Chemla, \textit{Nature} \textbf{417}, 47 (2002).

\bibitem{timofeev1}
A. V. Larionov, V. B. Timofeev, P. A. Ni, S. V. Dubonos, I. Hvam, K. Soerensen, JETP Lett. \textbf{75}, 570 (2002);
V. B. Timofeev, Phys. Usp. \textbf{48}, 295-306 (2005).

\bibitem{jetpl0840329}
A. V. Gorbunov, V. B. Timofeev, JETP Lett. \textbf{84}, 329 (2006); M. Alloing, M. Beian, M. Lewenstein, D. Fuster, Y. Gonz\'alez, L. Gonz\'alez,
R. Combescot, M. Combescot, and F. Dubin, Europhys. Lett. \textbf{107}, 10012 (2014).

\bibitem{butov2_}
A. A. High, J. R. Leonard, A. T. Hammack, M. M. Fogler, L. V. Butov, A. V. Kavokin, K. L. Campman, and A. C. Gossard, \textit{Nature} \textbf{483}, 584 (2012).

\bibitem{jetpl0740288}
Yu. E. Lozovik and I. V. Ovchinnikov, JETP Letters \textbf{74}, 288 (2001);
Yu. E. Lozovik and I. V. Ovchinnikov, Phys. Rev. B \textbf{66}, 075124 (2002).

\bibitem{prb091161413}
C.-E. Bardyn, T. Karzig, G. Refael, and T. C. H. Liew, Phys. Rev. B \textbf{91}, 161413(R) (2015).

\bibitem{prb082014508}
I. L. Kurbakov, Yu. E. Lozovik, G. E. Astrakharchik, and J. Boronat, Phys. Rev. B \textbf{82}, 014508 (2010).

\bibitem{prb088195309}
A. V. Kavokin, M. Vladimirova, B. Jouault, T. C. H. Liew, J. R. Leonard, L. V. Butov, Phys. Rev. B \textbf{88}, 195309 (2013).

\bibitem{prb089035302}
H. Sigurdsson, T. C. H. Liew, O. Kyriienko, and I. A. Shelykh, Phys. Rev. B \textbf{89}, 035302 (2014).

\bibitem{prl110246403}
A. A. High, A. T. Hammack, J. R. Leonard, Sen Yang, L.V. Butov, T. Ostatnicky, M. Vladimirova, A. V. Kavokin, T. C. H. Liew, K. L. Campman, and A. C. Gossard, Phys. Rev. Lett. \textbf{110}, 246403 (2013).

\bibitem{kukushkin}
V. V. Soloviev, I. V. Kukushkin, J. Smet, K. von Klitzing, W. Dietsche, JETP Letters \textbf{83}, 553 (2006).

\bibitem{purcell}
E. M. Purcell, Phys. Rev. \textbf{69}, 681 (1946).

\bibitem{kleppner}
D. Kleppner, Phys. Rev. Lett. \textbf{47}, 233 (1981);
E. Yablonovitch, Phys. Rev. Lett. \textbf{58}, 2059 (1987).

\bibitem{bayer}
Y. Yamamoto, S. Machida, G. Bj\"{o}rk, Phys. Rev. A \textbf{44}, 657 (1991);
M. Bayer, F. Weidner, A. Larionov, A. McDonald, A. Forchel, T. L. Reinecke, Phys. Rev. Lett. \textbf{86}, 3168 (2001).

\bibitem{noda}
M. Fujita, S. Takahashi, Y. Tanaka, T. Asano, S. Noda, \textit{Science} \textbf{308}, 1296 (2005); A. Kress, F. Hofbauer, N. Reinelt, M. Kaniber, H. J. Krenner, R. Meyer, G. B\"{o}hm, and J. J. Finley, Phys. Rev. B \textbf{71}, 241304 (2005); D. Englund, D. Fattal, E. Waks, G. Solomon, B. Zhang, T. Nakaoka, Y. Arakawa, Y. Yamamoto, and J. Vu\v{c}kovi\'{c}, Phys. Rev. Lett. \textbf{95}, 013904 (2005).

\bibitem{Popov_}
V. N. Popov, {\it Functional Integrals in Quantum Field Theory and Statistical Physics} (Reidel, Dordrecht, 1983).

\bibitem{kane_}
J. W. Kane, L. P. Kadanoff, Phys. Rev. \textbf{155}, 80 (1967).

\bibitem{minnhagen_}
P. Minnhagen and G. G. Warren, Phys. Rev. B {\bf 24}, 2526 (1981).

\bibitem{giorgini_}
S. Giorgini, L. Pitaevskii, and S. Stringari, Phys. Rev. B \textbf{49}, 12938 (1994).

\bibitem{pump} A. T. Hammack, M. Griswold, L.V. Butov, L. E. Smallwood, A. L. Ivanov, and A. C. Gossard, Phys. Rev. Lett. \textbf{96}, 227402 (2006).

\bibitem{prb059001625_}
L. V. Butov, A. Imamoglu, A. V. Mintsev, K. L. Campman and A. C. Gossard, Phys. Rev. B \textbf{59}, 1625 (1999).

\bibitem{prb045011403_}
R. Eccleston, B. F. Feuerbacher, J. Kuhl, W. W. R\"{u}hle, and K. Ploog, Phys. Rev. B \textbf{45}, 11403 (1992).

\bibitem{prb050014416}
S.-F. Ren, J.-B. Xia, H.-X. Han, and Z.-P. Wang, Phys. Rev. B \textbf{50}, 14416 (1994).

\bibitem{prb042008928}
2D exciton Bohr radius $a_B^X$ is known to be 2 times smaller than in 3D, its binding energy $E_b$ and $1/(a_B^X)^2$ are 4 times higher. Therefore for an ultra-narrow QW, densities can be assumed four times as high as in a very wide (27-nm) QW \cite{prb045011403_}, where $E_b$ and $1/(a_B^X)^2$ are almost the same as in bulk. See L. C. Andreani, A. Pasquarello, Phys. Rev. B \textbf{42}, 8928 (1990).

\bibitem{cqw_mott_tr}
G. Kir\v{s}ansk\.{e}, P. Tighineanu, R. S. Daveau, J. Miguel-S\'{a}nchez, P. Lodahl, and Soren St{\o}bbe, Phys. Rev. B \textbf{94}, 155438 (2016).

\bibitem{tbp}
N.S. Voronova, I.L. Kurbakov, A.S. Pliashechnik, and Yu.E. Lozovik (to be published).

\bibitem{hohenberg}
P. C. Hohenberg, Phys. Rev. \textbf{158}, 383 (1967).

\bibitem{jpc006001181_}
J. M. Kosterlitz, D. J. Thouless, J. Phys. C {\bf 6}, 1181 (1973).

\bibitem{hadzibabic}
Z. Hadzibabic, P. Kr\"{u}ger, M. Cheneau, B. Battelier, J. Dalibard, \textit{Nature} \textbf{441}, 1118 (2006).

\bibitem{jpc007001046_}
J. M. Kosterlitz, J. Phys. C {\bf 7}, 1046 (1974).

\bibitem{prl040000783_}
V. Ambegaokar, B. I. Halperin, D. R. Nelson, and E. D. Siggia, Phys. Rev.
Lett. {\bf 40}, 783 (1978).

\bibitem{SLY}
Yu. E. Lozovik, V. I. Yudson, Physica A {\bf 93}, 493 (1978).

\bibitem{prl102180404}
C. Mora and Y. Castin, Phys. Rev. Lett. {\bf 102}, 180404 (2009).

\bibitem{pra081013612_}
G. E. Astrakharchik, J. Boronat, I. L. Kurbakov, Yu. E. Lozovik, and F. Mazzanti, Phys. Rev. A {\bf 81}, 013612
(2010).

\bibitem{prl039001201_}
D. R. Nelson and J. M. Kosterlitz, Phys. Rev. Lett. {\bf 39}, 1201 (1977).

\bibitem{pla366000487_}
Yu. E. Lozovik, I. L. Kurbakov, and M. Willander, Phys. Lett. A {\bf 366}, 487 (2007).

\bibitem{exchange}
For direct excitons, exchange energy is of the order of $0.01$~meV (see H. Fu, L.-W. Wang, and A. Zunger, Phys. Rev. B {\bf 59}, 5568 (1999)), which corresponds to the magnetic field of the order of $0.1$~T.

\bibitem{prb085045207}
K. Sivalertporn, L. Mouchliadis, A. L. Ivanov, R. Philp, and E. A. Muljarov, Phys. Rev. B \textbf{85}, 045207 (2012).

\bibitem{butov3}
M. M. Fogler, L. V. Butov, and K. S. Novoselov, Nat. Comm. \textbf{5}, 4555 (2014).

\bibitem{glazov}
M. Kulig, J. Zipfel, P. Nagler, S. Blanter, C. Sch\"{u}ller, T. Korn, N. Paradiso, M. M. Glazov, and A. Chernikov, Phys. Rev. Lett. \textbf{120}, 207401 (2018).

\end{thebibliography}

\begin{thebibliography}{199}
\bibitem{prb059001625}
L. V. Butov, A. Imamoglu, A. V. Mintsev, K. L. Campman and A. C. Gossard, Phys. Rev. B \textbf{59}, 1625 (1999).

\bibitem{prb045011403}
R. Eccleston, B. F. Feuerbacher, J. Kuhl, W. W. R\"{u}hle, and K. Ploog, Phys. Rev. B \textbf{45}, 11403 (1992).

\bibitem{kukushkin1}
V. V. Soloviev, I. V. Kukushkin, J. Smet, K. von Klitzing, W. Dietsche, JETP Letters \textbf{84}, 222 (2006).

\bibitem{kane}
J. W. Kane and L. P. Kadanoff, Phys. Rev. \textbf{155}, 80 (1967).

\bibitem{kubo}
R. Kubo, J. Phys. Soc. Japan \textbf{17}, 1100 (1962).

\bibitem{rmp040000411}
P. Carruthers and M. M. Nieto, Rev. Mod. Phys. {\bf 40}, 411 (1968).

\bibitem{jpc007001046}
J. M. Kosterlitz, J. Phys. C {\bf 7}, 1046 (1974).

\bibitem{prb024002526}
P. Minnhagen and G. G. Warren, Phys. Rev. B {\bf 24}, 2526 (1981).

\bibitem{prl040000783}
V. Ambegaokar, B. I. Halperin, D. R. Nelson, and E. D. Siggia, Phys. Rev.
Lett. {\bf 40}, 783 (1978).

\bibitem{prb049001205}
H.-F. Meng, Phys. Rev. B {\bf 49}, 1205 (1994).

\bibitem{prb049012938}
S. Giorgini, L. Pitaevskii, and S. Stringari, Phys. Rev. B \textbf{49}, 12938 (1994).

\bibitem{Popov}
V. N. Popov, {\it Functional Integrals in Quantum Field Theory and Statistical Physics} (Reidel, Dordrecht, 1983).

\bibitem{jpc006001181}
J. M. Kosterlitz and D. J. Thouless, J. Phys. C {\bf 6}, 1181 (1973).

\bibitem{prl039001201}
D. R. Nelson and J. M. Kosterlitz, Phys. Rev. Lett. {\bf 39}, 1201 (1977).

\bibitem{pla366000487}
Yu. E. Lozovik, I. L. Kurbakov, and M. Willander, Phys. Lett. A {\bf 366}, 487 (2007).

\bibitem{butov2}
A. A. High, J. R. Leonard, A. T. Hammack, M. M. Fogler, L. V. Butov, A. V. Kavokin, K. L. Campman, and A. C. Gossard, \textit{Nature} \textbf{483}, 584 (2012).

\end{thebibliography}
\end{document}